# INCREASING COMPRESSION RATIO OF LOW COMPLEXITY COMPRESSIVE SENSING VIDEO ENCODER WITH APPLICATION-AWARE CONFIGURABLE MECHANISM

*Shuang Yu[2], Fei Qiao[1], Li Luo[2] and Huazhong Yang[1]*

National Laboratory for Information Science and Technology,
Dept. of Electronic Engineering, Tsinghua University, Beijing 100084, P.R. China[1]
Department of Electronics, Beijing Jiaotong University[2]

## ABSTRACT

With the development of embedded video acquisition nodes and wireless video surveillance systems, traditional video coding methods could not meet the needs of less computing complexity any more, as well as the urgent power consumption. So, a low-complexity compressive sensing video encoder framework with application-aware configurable mechanism is proposed in this paper, where novel encoding methods are exploited based on the practical purposes of the real applications to reduce the coding complexity effectively and improve the compression ratio (CR). Moreover, the group of processing (GOP) size and the measurement matrix size can be configured on the encoder side according to the post-analysis requirements of an application example of object tracking to increase the CR of encoder as best as possible. Simulations show the proposed framework of encoder could achieve 60X of CR when the tracking successful rate (SR) is still keeping above 90%.

## 1. INTRODUCTION

In current video surveillance system, traditional video coding methods, such as H.264, MPEG, are exploited to compress the signal at video acquisition node. However, with the development of low power video acquisition node and wireless video surveillance network, in which the power, storage and processing ability are limited, the traditional video coding method is not suitable any more due to its high complexity. So many attempts have been done to develop low complexity video coding method to meet the application requirement. The methods to improve the conventional video coder focus on reducing the computation complexity [1-2], but the effect is not significant. There are also some novel video coding methods emerging, such as Distributed Video Coding (DVC) [3], Blind Signal Separation (BSS) [4] and Compressive Sensing(CS) [5-6] which though they have low complexity, but they are not able to achieve high compression. However, the method using compressive sensing is more suitable for video surveillance according to the special characteristics of surveillance video that has a well defined, relatively static background to make high correlation of successive frames.

There have been an abundance of research activities in video surveillance using compressive sensing. [7] has proposed the application of compressive sensing in wireless video surveillance system that encodes the video directly according to the theory of compressive sensing leading to a low compressing efficiency. Video tracking algorithms using compressive sensing were presented in [8-9]. It only pays attention to improving tracking algorithm, but doesn't solve the problem of encoder. In [10-11], new coding schemes which are suitable for surveillance applications allow tracking and detecting video objects without the need to recover the video have been put forward. Particularly, they employed compressive sensing for object detection which only reconstruct foreground image on account of background subtraction [12-13]. Though they can lead to an efficient coding method to meet the designated applications, yet they are not able to achieve high compression and can't get complete video for other surveillance applications.

In this paper, we propose a video coding framework based on compressive sensing according to the characteristics of surveillance video in which the encoder can achieve not only very low complexity but also high compression. And the video after recovery can be effectively used in many aspects of the video surveillance. We verify the effectiveness using target tracking in this paper. In the proposed framework, multiple-residual method is employed to improve the compression efficiency according to the characteristics of video surveillance system that not each frame needs to have high resolution. The source video is divided into key frame and several CS frames to encode. Traditional video coding method is employed to encode key frame that can keep the quality of recovery. Gaussian random matrix is used to make measurements of CS frames. As the same time, we can configure the GOP size and the measurement matrix size according to the different applications. Thus, the CR can be increased as best as possible.

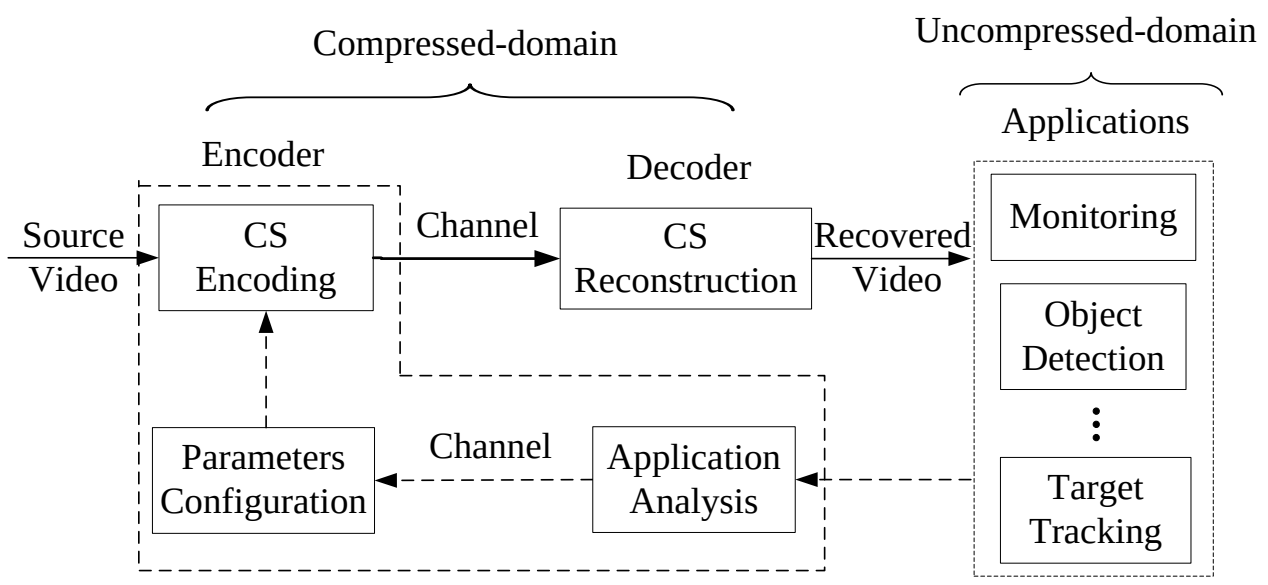

Fig.1 Video coding framework for video surveillance

The rest of the paper is organized as follows. In section II, video coding framework and the special characteristics of video surveillance system are stated. In section III, the improved multiple residual video coding method using compressive sensing is described in detail. Then the simulation results to validate this method are presented in section IV. Finally, the conclusion and discussion are given in section V.

## 2. VIDEO CODING FRAMEWORK FOR VIDEO SURVEILLANCE

In this section, we propose a framework in which the video source is coded using compressive sensing based on the characteristic of video surveillance. After encoding, the video stream is broadcast to decoder. And the video after reconstruction is used in many aspects of video surveillance system. Particularly, we can configure the GOP size and the measurement matrix size of encoder according to the different applications' requirement. The framework is showed in Fig.1.

Our goal is to develop a video coding method that can not only solve the issue of complexity and communication bandwidth but also meet the demand of video surveillance. So we can combine the encoding approach and surveillance applications to improve the performance of encoder.

### 2.1. The characteristics of video surveillance

For surveillance video, it exists a well defined, relatively static background, so the successive frames have very high correlation. And with the widely application of video surveillance, it is always applied in resource-constrained environments that the cameras transmit surveillance video to a processing centre where the videos are processed and analyzed. So we don't want to have a lot preprocessing to increase coding complexity at video acquisition node.

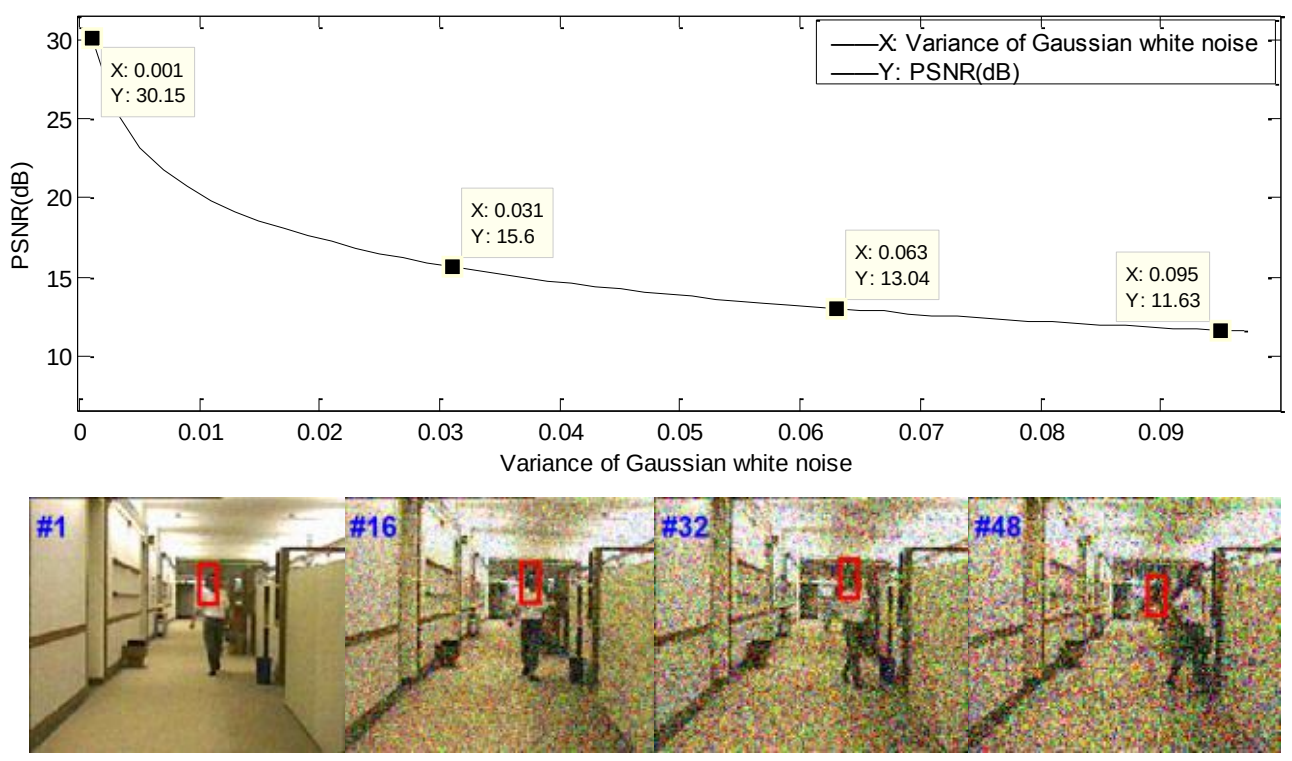

Fig.2 Upper curve is the PSNR with the change of Gaussian noise variance. Under pictures are the corresponding points of the tracking results (#1, #16, #32 frames are tracking successfully, and #48 frame is failed).

Further, in the video surveillance system, the video after recovery can be used in monitoring, moving target detection, object recognition, target tracking and so on. Not each frame needs to have high resolution according to the features of these applications. When the video quality decreases progressively within a certain range, it can still be used to complete these applications. We verify this character using compressive tracking algorithm by adding Gaussian noise to decrease the quality of source video gradually. The Fig.2 shows the special characteristic. We can see that it almost has no effect on target tracking in a certain range. But when the video quality drops to a certain degree, errors encounter.

### 2.2. The design of video coding method

According to the characteristic of video surveillance, the residual method based on compressive sensing is more suitable for the application of video surveillance in which the residual of adjacent frames is quite sparse that helps to reduce the number of measurements to improve compression rate, and has low computational complexity. Furthermore, this method processes each frame of the video respectively that more conductive to the requirement of storage and real-time. But the existing residual methods based on compressive sensing employ a key frame and a CS frame. Although these two frames all have high recovery quality, the compression efficiency is so poor that no advantage exists in the communication bandwidth and transmission power. So we need to improve this method making it more efficient on video surveillance. Moreover, the encoder structure is not fixed, but can be configured by the demand of surveillance applications.

## 3. MULTIPLE RESIDUAL CODING METHOD USING COMPRESSIVE SENSING

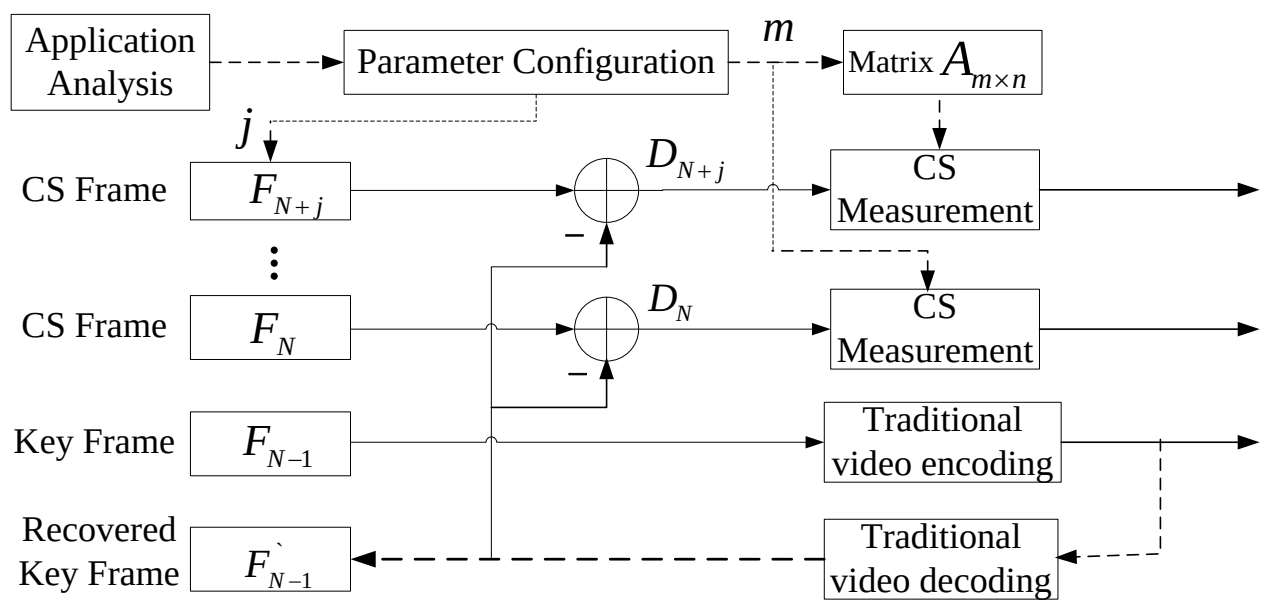

Fig.3 Video encoding using compressive sensing

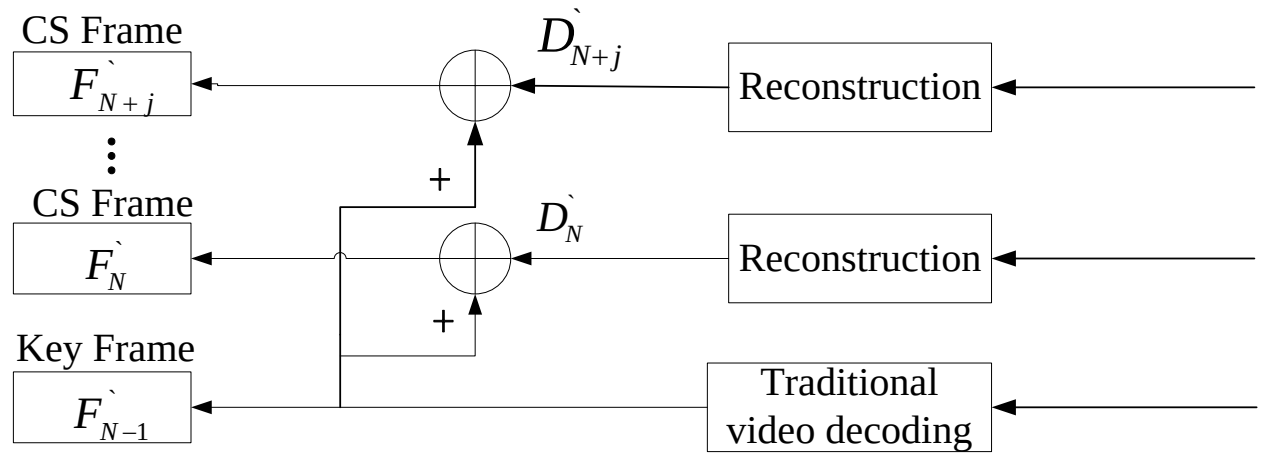

Fig.4 Video decoding using compressive sensing

The multiple-residual video encoding method is proposed in this section that the source video is firstly divided into key frame and several residual frames to encode. The key frame employs traditional video coding method to ensure recovery quality and the residual frames use CS coding method to simplify complexity. After recovery, the video is well used in surveillance applications. Meanwhile, the GOP size and measurement matrix size can be configured by parameters configurator according to the application analyzer.

### 3.1. Video encoding using compressive sensing

A source video consists of a number of frames with a resolution of $M \times N$, where M and N are the numbers of the horizontal and vertical pixels, respectively, for each frame of the video. To efficiently encode it, we divide the source video into a key frame and several residual frames. Due to the high correlation of successive surveillance video frames, the residual frames are so sparse that can achieve high compression rate using compressive sensing. Meanwhile, they have low complexity. The key frame is crucial to the video coding because it is the basis for decoding the CS frames. The CS frames can be decoded only after the key frame is recovered successfully. So we adopt the traditional video coding method to ensure the recovery quality. The framework is illustrated in Fig.3.
The encoder adopts random matrix to measure each CS frame of the video respectively. It has more advantages on storage and real-time. Particularly, the CR of encoder is configurable due to the change of GOP size and measurement matrix size.

### 3.2. Video decoding using compressive sensing

There are two steps for recovering the video. First, we reconstruct the key frame using traditional video decoding method. Then, the recovered key frame is reserved to reconstruct the other CS frames.
Because of the sparsity of the residual, we can reconstruct the residual successfully by solving the constraint minimization problem:

$$\min_x \Phi(x), s.t. Ax = y, \qquad (1)$$

where $x \in R^{M\times N}$ with $M \times N = n$ is the original sparse signal, $A \in R^{m\times n} (m < n)$ is the measurement matrix, and $y \in R^m$ is the measurements. $\Phi(x)$ is the regularization term to handle the ill-posedness or to prohibit overfitting. After the reconstruction, we add the residual to the previous key frame to recover the CS frames. Fig.4 shows the decoding scheme.

In this framework, we employ multiple residual methods to improve compression performance. The CS frames all subtract the same key frame in a GOP. Although the quality gradually reduces at the same compression rate, but it almost has no effect on surveillance applications. So we can increase the number of CS frames as far as possible in the premise of completing surveillance applications. Thus, we can achieve high compression performance. Moreover, we reconstruct key frame, which is subtracted from CS frames to acquire residual, rather than to subtract the original key frame. This method can remove coding and decoding errors of the key frame. The function can be described as below:

$$F^{`}_{N-1} = F_{N-1} + e_{coding} + e_{decoding} \qquad (2)$$

$$D^{`}_N = F_N - F^{`}_{N-1} = F_N - F_{N-1} - e_{coding} - e_{decoding} + e \qquad (3)$$

$$F^{`}_N = D^{`}_N + F^{`}_{N-1} = F_N + e \qquad (4)$$

where $e_{coding}$ and $e_{decoding}$ are the encoding and decoding errors of key frame. $e$ is the errors of CS frames.

## 4. SIMULATION

Simulations are performed on *hall monitor* and *biker*. It is QCIF resolution (176*144) with a frame rate of 30 f/s. In the simulation, the source video is encoded using compressive sensing as described in section III. The framework uses a key frame and four CS frames. A Gaussian random matrix is used to make the measurements. The compression ratio of all CS frames is 50. And the video reconstruction is performed by TVAL3 [14] algorithm. The key frame employs H.264 intra

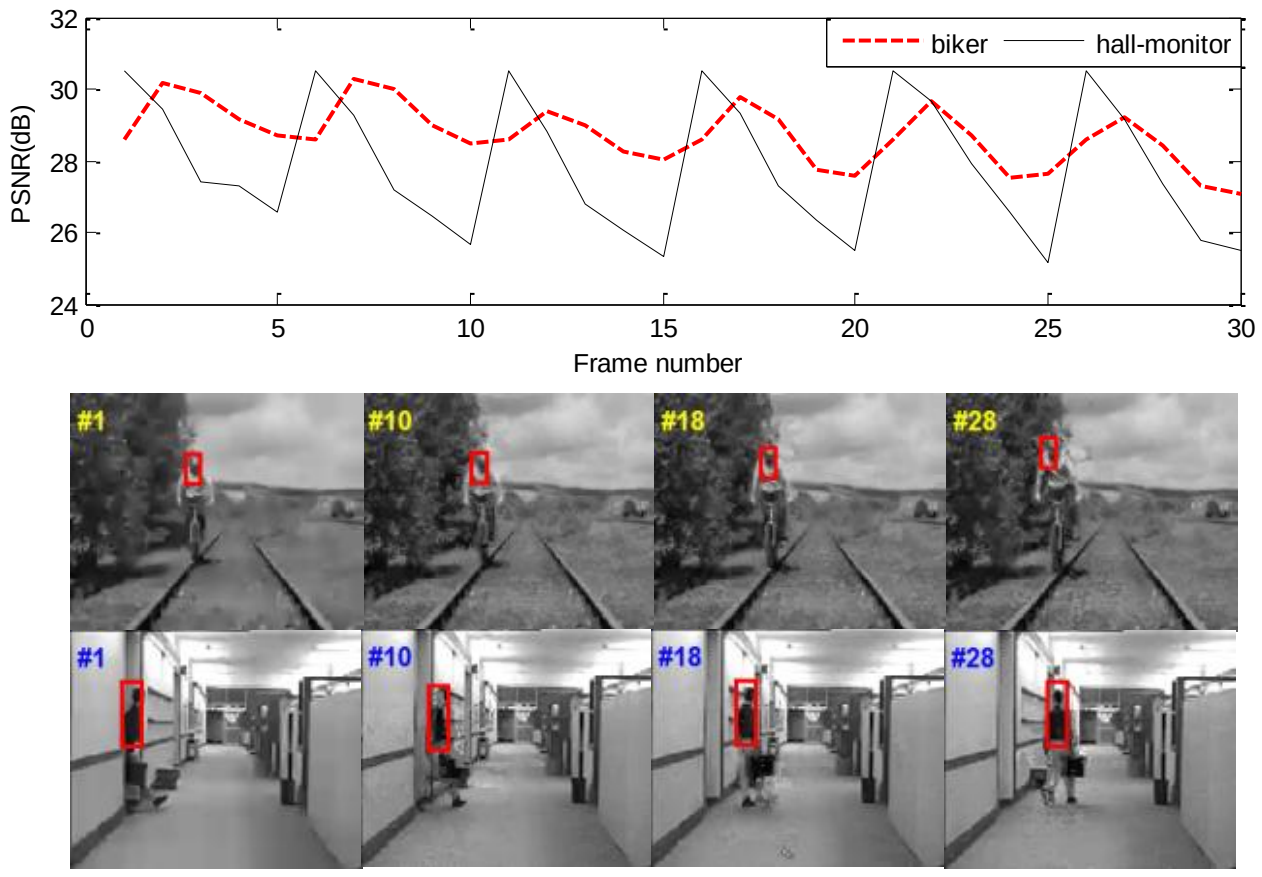

Fig.5 The PSNR of proposed framework and tracking results of recovered video

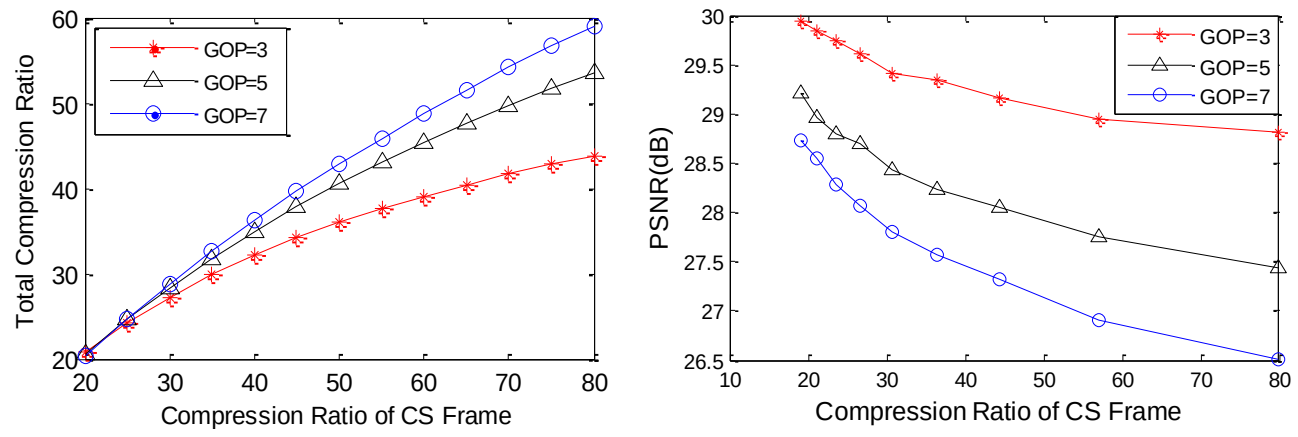

(a) (b)

Fig.6 (a) is the change of total compression ratio with the compressive ratio of CS frames. (b) is the change of PSNR with the compressive ratio of CS frames.

and the compression rate is 23. Fig.5 is the reconstruction PSNR and tracking results. We can see the key frame has high quality and the CS frame quality decreases frame by frame. It accords with the design of encoder. After recovery, the video is used to track target using tracking algorithm. The result demonstrates that it can track the target accurately. It proves the effectiveness of proposed framework.

From the Fig.6, we can configure the GOP size and CR of CS frames according to the different applications. They affect the total CR and PSNR of the encoder. Thus, the proposed framework can not achieve low complexity but achieve an increasing compression ratio.

The table I sums up the influence of GOP size and the rate of key frame' CR and CS frame' CR on compression efficiency, tracking successful rate (SR) and PSNR. We can see when GOP size is fixed, the tracking SR decreases with the increasing of compression ratio. And when compression ratio is fixed, the SR decreases with the increasing GOP size. So we can improve compression ratio properly to meet the requirement of application. It fully shows the advantages of framework proposed in this paper.

Table I Compression efficiency, complexity and track rate

| Total CR | PSNR (dB) | Track SR | GOP | rate |
|---|---|---|---|---|
| 32.09 | 29.21 | 99% | 3 | 23:40 |
| 39.06 | 28.95 | 98% | 3 | 23:60 |
| 43.81 | 28.81 | 96% | 3 | 23:80 |
| 34.85 | 28.14 | 99% | 5 | 23:40 |
| 45.39 | 27.75 | 95% | 5 | 23:60 |
| 53.49 | 27.43 | 92% | 5 | 23:80 |
| 36.18 | 27.43 | 99% | 7 | 23:40 |
| 48.79 | 29.91 | 95% | 7 | 23:60 |
| 59.08 | 26.51 | 90% | 7 | 23:80 |

## 5. CONCLUSION

This paper starts with analyzing the existing video coding methods using compressive sensing and the characteristics of video surveillance. Combining with the both advantages, we propose multiple-residual video coding framework based on compressive sensing according to the characteristics of video surveillance system in which the encoder can achieve not only very low complexity but also high compression. And the video after recovery can be effectively used in many aspects of video surveillance. Moreover, the GOP size and compression rate of the encoder can be configured according to the requirement of various applications. Simulations demonstrate that the proposed method has higher compression efficiency and lower complexity. Real-time compressive tracking algorithm is used to verify the effectiveness of the system.